\documentclass{aa}
\usepackage[varg]{txfonts}
\usepackage{wasysym}
\usepackage{natbib}
\usepackage{adjustbox}
\usepackage{graphicx}
\usepackage{dblfloatfix}
\graphicspath{ {images/} }
\bibpunct{(}{)}{;}{a}{}{,}
\begin{document}
\title{$^{15}$N Fractionation in Infrared-Dark Cloud Cores}

\author{S. Zeng\inst{1}, I. Jim\'enez-Serra\inst{1}, G. Cosentino\inst{2}, S. Viti\inst{2}, A. T. Barnes\inst{3,4}, J. D. Henshaw\inst{3},  P. Caselli\inst{4}, F. Fontani\inst{5}, P. Hily-Blant\inst{6}
}

\institute{School of Physics and Astronomy, Queen Mary University of London,
Mile End Road, E1 4NS London, United Kingdom
\and
University College London, 132 Hampstead Road, London, NW1 2PS, UK
\and 
Astrophysics Research Institute, Liverpool John Moores University, Liverpool, L3 5RF, UK
\and
Max-Planck Institute for Extraterrestrial Physics, Giessenbachstrasse 1, 85748 Garching, Germany
\and 
INAF-Osservatorio Astrofisico di Arcetri, Largo E. Fermi 5, 50125, Firenze, Italy
\and
Institut de Plan\'etologie et d'Astrophysique de Grenoble, 414 rue de la Piscine, F-38041 Grenoble, France}

\date{Received date /
Accepted date }

\abstract
{Nitrogen is one of the most abundant elements in the Universe and its $^{14}$N/$^{15}$N isotopic ratio has the potential to provide information about the initial environment in which our Sun formed. Recent findings suggest that the Solar System may have formed in a massive cluster since the presence of short-lived radioisotopes in meteorites can only be explained by the influence of a supernova.}
{To determine the $^{14}N$/$^{15}N$ ratio towards a sample of cold, massive dense cores at the initial stages in their evolution.}
{We have observed the J=1$\rightarrow$0 transitions of HCN, H$^{13}$CN, HC$^{15}$N, HN$^{13}$C and H$^{15}$NC toward a sample of 22 cores in 4 Infrared-Dark Clouds (IRDCs). IRDCs are believed to be the precursors of high-mass stars and star clusters. Assuming LTE and a temperature of 15$\,$K, the column densities of HCN, H$^{13}$CN, HC$^{15}$N, HN$^{13}$C and H$^{15}$NC are calculated and their $^{14}N$/$^{15}N$ ratio is determined for each core.}
{The $^{14}N$/$^{15}N$ ratio measured in our sample of IRDC cores range between $\sim$70 and $\geq$763 in HCN and between $\sim$161 and $\sim$541 in HNC. They are consistent with the terrestrial atmosphere (TA) and protosolar nebula (PSN) values, and with the ratios measured in low-mass pre-stellar cores. However, the $^{14}N$/$^{15}N$ ratios measured in cores C1, C3, F1, F2 and G2 do not agree with the results from similar studies toward the same massive cores using nitrogen bearing molecules with nitrile functional group (-CN) and nitrogen hydrides (-NH) although the ratio spread covers a similar range.}
{Amongst the 4 IRDCs we measured relatively low $^{14}N$/$^{15}N$ ratios towards IRDC G which are comparable to those measured in small cosmomaterials and protoplanetary disks. The low average gas density of this cloud suggests that the gas density, rather than the gas temperature, may be the dominant parameter influencing the initial nitrogen isotopic composition in young PSN.}

\keywords{ISM: molecules -- Astrochemistry -- Stars: formation}
\titlerunning{$^{15}$N Fractionation in Infrared-Dark Cloud Cores}
\authorrunning{S.Zeng et al.}
\maketitle


\section{Introduction}

{For decades, our Solar System was believed to have resided in a relatively isolated, low-mass molecular cloud core during its formation. However the detection of short-lived radioactive species in meteorites have suggested a different scenario in which the birthplace of the Sun may have been a massive cluster affected by a supernova event \citep{Adams2010,Dukes2012,Pfalzner2013,Nicholson2017}. If so, the initial chemical composition of our Solar System, and thus of planets, meteorites and comets, may have been affected by the same physical process.}

{Measurements of the abundance isotopic ratios of the elements can be used to unveil the initial chemical composition of the proto-solar nebulae (PSN) from which the Solar System formed. The isotopic ratios of carbon ($^{12}$C/$^{13}$C) and oxygen ($^{16}$O/$^{18}$O) show a remarkable agreement among cometary materials, the local interstellar medium (ISM) and the Solar value \citep{Manfroid2009,Milam2005,Wilson1994}. Nitrogen, by contrast, has a peculiar behaviour since its $^{14}$N/$^{15}$N isotopic ratio exhibits discrepancies across various environments within the Solar System. The $^{14}$N/$^{15}$N ratio measured in Jupiter's atmosphere \citep[450$\pm$100,][]{Fouchet2004} is considered as the most representative value of the PSN and it matches the present day solar wind value \citep[441$\pm$6,][]{Marty2010}. However, the ratios measured in the terrestrial atmosphere (TA) (\citealp[$\sim$272 in Earth,][]{Junk1958}; \citealp[272$\pm$54 in Venus,][]{Hoffman1979}; \citealp[173$\pm$11 in Mars,][]{Wong2013}), comets (\citealp[147.8$\pm$5.7,][]{Manfroid2009}, \citealp[139$\pm$26 from HCN and 165$\pm$40 from CN,][]{Bockelee-Morvan2008}), Interplanetary Dust Particles \citep[or IDPs; values of 180-305;][]{Floss2006} and meteorites \citep[192-291,][]{Alexander2007}, are lower than that measured in Jupiter's atmosphere.}

{In molecular clouds, the discrepancies in the $^{14}$N/$^{15}$N isotopic ratios spread over a larger range. In contrast to many molecular species (e.g. CO), N-bearing molecules do not suffer significant freeze-out onto grains in the coldest, densest regions of IRDCs, and are therefore reliable tracers of the gas chemistry and kinematics in cores. The nitrogen fractionation mechanisms are either due to chemical fractionation \citep{Terzieva2000,Rodgers2008,Wirstrom2012,Hily-Blant2013a} or selective photodissociation effect \citep{Lyons2009,Heays2014}. IRDCs are dense and highly extinguished \citep[with visual extinctions $>$10$\,$mag;][]{Kainulainen2013}, and therefore selective photodissociation is not expected to play an important role since this process becomes inefficient at A$_v$ $\geq$ 3$\,$mag \citep{Heays2014}. As for chemical fractionation, these tracers can be categorized into: 1) Hydride-bearing molecules with an amine (-NH) functional group believed to have originated from reactions with N$_2$; and 2) Nitrile-bearing molecules with a nitrile (-CN) functional group that form via reactions with atomic N \citep{Rodgers2008, Hily-Blant2013a}. Numerous measurements of the $^{14}$N/$^{15}$N ratio exist towards low-mass pre-stellar cores \citep[334$\pm$50, 1000$\pm$200 and 230$\pm$90 from NH$_3$, N$_2$H$^+$ and HCN, respectively;][]{Lis2010,Bizzocchi2013,Hily-Blant2013a} and protostars \citep[$\sim$160-290 from HCN and HNC;][]{Wampfler2014}, but observations of this ratio towards their massive counterparts are lacking.}

{A re-investigation of the fractionation processes of nitrogen in the ISM by \citet{Roueff2015} showed that nitrogen chemistry depends on the temperature and density of the primordial gas in the parental cloud. Since the Sun may have formed in a massive cluster, and since low-mass and high-mass star-forming regions present gas temperatures and densities that differ respectively by $\sim$5-10$\,$K and by factors of 10 \citep[][]{Pillai2006,Crapsi2007,Henshaw2013}, measurements of the $^{14}$N/$^{15}$N ratio in high-mass star-forming regions could provide insight into the initial bulk composition of the PSN.}

{Recently, \citet{Adande2012} and \citet{Fontani2015} have measured the $^{14}$N/$^{15}$N ratio towards a sample of high-mass star-forming regions. While in the \citet{Adande2012} sample the $^{14}$N/$^{15}$N ratios measured from CN and HNC lie between $\sim$120--400, in \citet{Fontani2015}'s work these measurements range from $\sim$180 to $\sim$1300 in N$_2$H$^+$ and $\sim$190 to $\sim$450 in CN. In both studies, the $^{14}$N/$^{15}$N ratios obtained from CN are comparable and fall between the TA and PSN values. However, only a few of these objects were pre-stellar in nature and larger samples of high-mass starless/pre-stellar cores are needed to measure the $^{14}$N/$^{15}$N isotopic ratio in regions with physical conditions resembling those of the early stages of the Solar System formation.}

{We present measurements of the $^{14}$N/$^{15}$N isotopic ratio in HCN and HNC obtained toward a sample of 22 high-mass cold cores embedded in 4 IRDCs. These cores are believed to represent the nurseries of high-mass stars and star clusters and have physical properties \citep[densities 10$^4$-10$^6$$\,$cm$^{-3}$ and temperatures $\leq$20$\,$K;][]{Pillai2006,Butler2012} similar to those expected for the initial conditions of the Solar System. In Section 2, we describe the observations and data analysis. The results are presented in Section 3.1 whilst the uncertainties involved in our calculations are discussed in Section 3.2.  In Sections 4.1, 4.2 and 4.3 we investigate the correlation of IRDC cores with star formation activity and compare our results with previous measurements of the $^{14}$N/$^{15}$N isotopic ratio in Solar System objects, low-mass and high-mass star-forming regions. In Sections 4.4 and 4.5, we discuss the effects of $^{13}$C depletion on the chemistry of nitrogen fractionation and the systematic trend observed in young and quiescent IRDCs with respect to more evolved, star-forming IRDCs. Our conclusions are presented in Section 5.}

\section{Observations}

{Observations of the J=1$\rightarrow$0 rotational transition of HCN, H$^{13}$CN, HC$^{15}$N, HN$^{13}$C and H$^{15}$NC were obtained with the IRAM-30m telescope\footnote{Based on observations carried out under projects number 134-12 and 027-13 with the IRAM 30m Telescope. IRAM is supported by INSU/CNRS (France), MPG (Germany) and IGN (Spain).} towards 22 massive cores embedded in IRDCs G028.37+00.07, G034.43+00.24, G034.77-00.55 and G035.39-00.33 \citep[hereafter Clouds C, F, G and H respectively, as in][]{Butler2012}. The frequencies of the transitions and the molecular data are included in Table~\ref{lines} whereas the properties of each IRDC are listed in Table~\ref{IRDCprop}. The EMIR receivers were tuned at 87$\,$GHz and the FTS spectrometer provided a spectral resolution of 200$\,$kHz (or $\sim$0.68$\,$kms$^{-1}$). We note that the HNC(J=1$\rightarrow$0) transition was not covered within our frequency range. Typical system temperatures ranged from 106$\,$K to 199$\,$K. The half-power beam width (HPBW) of the telescope was 28$"$ at 87$\,$GHz. The spectra were measured in units of antenna temperature, T$^*_A$, and converted into main beam temperature, T$_{\rm mb}$, by using a beam efficiency of 0.81. Data reduction was carried out using the GILDAS/CLASS software package\footnote{See http://www.iram.fr/IRAMFR/GILDAS.}.
{The spectra of HCN (J=1$\rightarrow$0), H$^{13}$CN(J=1$\rightarrow$0), HC$^{15}$N(J=1$\rightarrow$0), HN$^{13}$C(J=1$\rightarrow$0) and H$^{15}$NC(J=1$\rightarrow$0) were obtained towards all cores reported by \citet{Butler2012} within Clouds F, G and H. For Cloud C, however, all cores were observed except C7 that laid outside our map. C7 will thus not be considered in our analysis.

\begin{table}[!tbp]
\caption{Observed HCN and HNC isotopologue transitions. Molecular data extracted from the JPL and CDMS molecular catalogues \citep{Pickett1998,Muller2005}. \label{lines}} 
\centering
\begin{adjustbox}{width=0.5\textwidth}
\begin{tabular}{cccccc}
\hline\hline
Molecules & Transition & Frequency [GHz] & A$_{\rm ul}$ [s$^{-1}$] & E$_{\rm u}$ [K] & g$_{\rm u}$ \\ 
\hline
H$^{13}$CN & J = 1$\rightarrow$0, F = 1$\rightarrow$1 & 86.33877 & 2.4$\times$10$^{-5}$ & 4.14 & 3 \\
H$^{13}$CN & J = 1$\rightarrow$0, F = 2$\rightarrow$1 & 86.34018 & 2.4$\times$10$^{-5}$ & 4.14 & 5 \\
H$^{13}$CN & J = 1$\rightarrow$0, F = 0$\rightarrow$1 & 86.34227 & 2.4$\times$10$^{-5}$ & 4.14 & 1 \\
HC$^{15}$N & J = 1$\rightarrow$0 & 86.05496 & 2.4$\times$10$^{-5}$ & 4.13 & 3 \\
HN$^{13}$C & J = 1$\rightarrow$0 & 87.09085 & 1.9$\times$10$^{-5}$ & 4.18 & 3 \\
H$^{15}$NC & J = 1$\rightarrow$0 & 88.86571 & 2.0$\times$10$^{-5}$ & 4.26 & 3 \\
HCN & J = 1$\rightarrow$0, F = 1$\rightarrow$1 & 88.63041 & 8.1$\times$10$^{-6}$ & 4.25 & 3\\
HCN & J = 1$\rightarrow$0, F = 2$\rightarrow$1 & 88.63160 & 8.1$\times$10$^{-6}$ & 4.25 & 5\\
HCN & J = 1$\rightarrow$0, F = 0$\rightarrow$1 & 88.63393 & 8.1$\times$10$^{-6}$ & 4.25 & 1\\
\hline
\end{tabular}
\end{adjustbox}
\end{table}

\begin{table*}[!tbp]
\tiny
\caption{Properties of IRDCs: galactic coordinates l and b, average peak radial velocity V$_{LSR}$, mass surface density $\Sigma$(sat), mass M in Solar mass \citep{Butler2012} and Galactocentric distance R$_{gc}$. \label{IRDCprop}}
\centering\begin{tabular}{cccccccc}
\hline\hline
IRDCs & l [$^{\circ}$] & b [$^{\circ}$] & V$_{LSR}$ [kms$^{-1}$] & $\Sigma$(sat) [gcm$^{-2}$] & M [M$_{\bigodot}$] & R$_{gc}$ [kpc] & $^{12}$C/$^{13}$C \\
\hline
C(G028.37+00.07) & 28.373 & 0.076 & 78.6 & 0.520 & 45000 & 4.65 & 40.2 \\
F(G034.43+00.24) & 34.437 & 0.245 & 57.1 & 0.370 & 4460 & 5.74 & 46.8\\
G(G034.77-00.55) & 34.771 & -0.557 & 43.5 & 0.347 & 2010 & 6.24 & 49.8\\
H(G035.39-00.33) & 35.395 & -0.336 & 44.7 & 0.416 & 13340 & 6.27 & 50.0\\
\hline
\end{tabular}
\end{table*}

\begin{figure}[!tbp]
\includegraphics[width=0.5\textwidth]{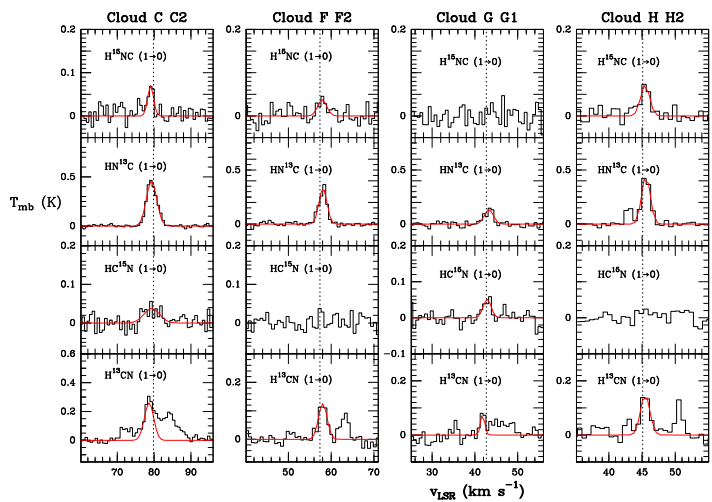}
\caption{\scriptsize From left to right, top to bottom panels: Emission lines of H$^{15}$NC, HN$^{13}$C, HC$^{15}$N, and hyperfine transitions of H$^{13}$CN observed with IRAM-30m toward IRDCs C, F, G and H. Red lines indicate the best Gaussian fit to the lines. Note that the hyperfine components of H$^{13}$CN were initially fitted but given the bad results of the fit, in a second step only the main $F$=2$\rightarrow$1 component of H$^{13}$CN was fitted using a single-Gaussian component profile (see red line in bottom panels and Section 3.1 for details). See also Figures A.1, A.2, A.3 and A.4 in Appendix for all spectra taken.}
\centering
\end{figure}

\section{Results}
\subsection{$^{14}$N/$^{15}$N ratios derived from the $^{13}$C isotopologues of HCN and HNC}

{In Fig.1, we show a sample of spectra of H$^{13}$CN, HC$^{15}$N, HN$^{13}$C and H$^{15}$NC obtained towards one massive core in each IRDC (the rest of spectra are shown in the Appendix in Figures A.1, A.2, A.3 and A.4). We assume that the emission from these isotopologues is optically thin and consider LTE conditions when calculating their column densities. For the excitation temperature T$_{\rm ex}$ of the molecular gas in these clouds, we assume a lower limit of 10K and an upper limit of 20K based on NH$_3$ measurements obtained toward other IRDCs \citep{Pillai2006}. Because the hyperfine structure of HN$^{13}$C cannot be resolved, we considered only one velocity component for this species, in the same way as for HC$^{15}$N and H$^{15}$NC. These lines were therefore fitted by using a single-component Gaussian profile. For H$^{13}$CN, the hyperfine structure of the $J$=1$\rightarrow$0 transition could be resolved in the spectra and the HFS line fitting method implemented in CLASS was initially used to obtain the optical depth of the H$^{13}$CN emission. We note however that the HFS fits presented large uncertainties (see the optical depth values in Table A.5) and therefore, in a second step, we fitted the main $F$=2$\rightarrow$1 component of H$^{13}$CN with a single-Gaussian component profile, calculated the column densities of H$^{13}$CN assuming optically thin emission, and corrected them by the statistical weight of the $F$=2$\rightarrow$1 transition. The measured integrated intensities, radial velocities, linewidths and peak intensities of the lines are listed in Appendix Tables A.1, A.2, A.3, and A.4. We consider that a line is detected when its peak intensity is $\geq$3$\sigma$, with $\sigma$ the rms noise level measured in the spectra (see column 2 in Tables A.1, A.2, A.3 and A.4). For the cores with no molecular detections, we used the 3$\sigma$ noise level as upper limits to their peak intensities. Note that if we take into account the optical depths derived from the HFS method (of $\sim$0.1-5.8), the resulting H$^{13}$CN column densities and hence the $^{14}$N/$^{15}$N ratio are increased by factors of 3--6. Nevertheless, they are consistent with the previously determined results within the uncertainties.}

{The corresponding $^{14}$N/$^{15}$N ratios were computed from the molecular column densities of the $^{14}$N and $^{15}$N HCN and HNC isotopologues after correcting by the $^{12}$C/$^{13}$C ratio for each IRDC. This ratio is calculated by using the Galactic $^{12}$C/$^{13}$C gradient as a function of Galactocentric distance derived by \citet{Milam2005} from CN measurements. The 
$^{12}$C/$^{13}$C ratios for IRDCs C, F, G and H are, respectively, 40.2, 46.8, 49.8 and 50.0 (see Table 2). The molecular column densities together with the derived $^{14}$N/$^{15}$N ratios at T$_{\rm ex}$=15$\,$K are listed in Table 3. 

In our measurements, the uncertainties in the $^{14}$N/$^{15}$N ratios were derived by propagating errors and by using the 1$\sigma$ uncertainties in the line integrated intensities calculated as rms $\times \sqrt{\Delta v \times \delta v}$, with $\Delta v$ the average linewidth of the line for all cores with emission and with $\delta v$ the velocity resolution of the spectrum ($\sim$0.68 km s$^{-1}$). The derived uncertainties for the $^{14}$N/$^{15}$N ratios are approximately $\sim$35$\%$. This may due to the weak detection of molecules in some of the cores. For the cores with no detections, the $^{14}$N/$^{15}$N ratios have been estimated using the 3$\sigma$ upper limits to the integrated intensities of H$^{15}$NC and HC$^{15}$N (see column 3 in Tables A.1, A.2, A.3 and A.4), and they should be considered as lower limits.

From Table~\ref{Ratios13C}, we find that the $^{14}$N/$^{15}$N ratios obtained towards Clouds C, F, G and H vary over a large range of values. In particular, for IRDCs C, F and H, the nitrogen ratios in HCN range between 122--$\geq$571, 282--$\geq$763, 142--458, respectively, while in HNC they range between $\geq$161-478 for Cloud C, 240--541 for Cloud F, and 234--488 for Cloud H. On the other hand, the $^{14}$N/$^{15}$N ratios measured toward the cores in Cloud G are systematically lower ranging between 70--$\geq$181 in HCN and 206--$\geq$237 in HNC. We have tested the effects of T$_{ex}$ on our results, and have found that if we use T$_{\rm ex}$=10$\,$K or T$_{\rm ex}$=20$\,$K instead of T$_{\rm ex}$=15$\,$K, the derived $^{14}$N/$^{15}$N isotopic ratios for both HCN and HNC do not vary significantly, lying within the $\sim$30\% uncertainties. Higher T$_{ex}$ (e.g. 50$\,$K and 100$\,$K) also confirm this behaviour, with the $^{14}$N/$^{15}$N ratios changing within a factor of 1.2. Nevertheless, \citet{Roueff2015} have recently pointed out that species such as HN$^{13}$C and H$^{13}$CN may suffer significant depletion in molecular clouds, challenging the interpretation of $^{14}$N/$^{15}$N isotopic ratios derived from $^{13}$C containing isotopologues. In Section 3.2, we explore this possibility by directly measuring the $^{14}$N/$^{15}$N ratios using HCN and its $^{15}$N isotopologue toward the IRDCs cores in our sample with optically thin HCN emission.

\begin{table}[!tbp]
\tiny
\caption{Column densities and nitrogen ratios obtained from the $^{13}$C isotopologues of HCN and HNC}
\label{Ratios13C}
\centering
\begin{adjustbox}{width=0.5\textwidth}
\begin{tabular}{ccccccc}
\hline\hline
 & H$^{13}$CN & HC$^{15}$N & HN$^{13}$C & H$^{15}$NC & HCN & HNC \\
\hline
Core & N$_{\rm tot}$(T$_{\rm ex}$=15K) & N$_{\rm tot}$(T$_{\rm ex}$=15K) & N$_{\rm tot}$(T$_{\rm ex}$=15K) & N$_{\rm tot}$(T$_{\rm ex}$=15K) & $^{14}$N/$^{15}$N & $^{14}$N/$^{15}$N  \\
 & [$\times$10$^{12}$cm$^{-2}$] & [$\times$10$^{11}$cm$^{-2}$] & [$\times$10$^{11}$cm$^{-2}$] & [$\times$10$^{11}$cm$^{-2}$] & (T$_{\rm ex}$=15K) & (T$_{\rm ex}$=15K) \\
\hline
\multicolumn{7}{c}{Cloud C} \\
\hline
C1 & 1.94 & $\leq$1.37 & 3.19 & 3.07 & $\geq$571$\pm$65 & 418$\pm$75 \\
C2 & 3.03 & 4.09 & 4.25 & 3.57 & 298$\pm$52 & 478$\pm$63\\
C3 & 0.89 & $\leq$2.1 & 0.98 & $\leq$2.45 & $\geq$170$\pm$64 & $\geq$161$\pm$14 \\
C4 & 2.02 & 4.20 & 3.42 & 5.07 & 193$\pm$30 & 271$\pm$31 \\
C5 & 2.34 & 2.80 & 3.02 & 3.54 & 337$\pm$89 & 342$\pm$40 \\
C6 & 1.92 & 2.26 & 2.44 & 2.34 & 343$\pm$83 & 420$\pm$82 \\
C7 & - & - & - & - & - & - \\
C8 & 1.23 & $\leq$1.33 & 1.94 & 2.41 & $\geq$371$\pm$60 & 325$\pm$73 \\
C9 & 2.74 & 9.01 & 3.94 & 7.59 & 122$\pm$20 & 209$\pm$31 \\
\hline
\multicolumn{7}{c}{Cloud F} \\
\hline
F1 & 1.82 & $\leq$1.12 & 2.05 & 3.99 & $\geq$763$\pm$82 & 240$\pm$26 \\
F2 & 1.15 & $\leq$1.25 & 2.05 & 2.53 & $\geq$431$\pm$63 & 378$\pm$77 \\
F3 & 1.11 & 1.51 & 2.25 & 1.95 & 346$\pm$80 & 541$\pm$79 \\
F4 & 1.83 & 3.03 & 2.78 & 3.47 & 282$\pm$60 & 374$\pm$57 \\
\hline
\multicolumn{7}{c}{Cloud G} \\
\hline
G1 & 0.35 & 2.47 & 0.92 & $\leq$1.94 & 70$\pm$28 & $\geq$237$\pm$21 \\
G2 & $\leq$0.31 & $\leq$1.54 & 1.71 & 4.14 & - & 206$\pm$31 \\
G3 & 0.63 & $\leq$1.72 & 1.74 & 3.65 & $\geq$181$\pm$54 & 237$\pm$36 \\
\hline
\multicolumn{7}{c}{Cloud H} \\
\hline
H1 & 0.98 & $\leq$1.39 & 1.91 & 4.09 & $\geq$353$\pm$51 & 234$\pm$20 \\
H2 & 0.97 & $\leq$1.32 & 1.90 & 3.11 & $\geq$366$\pm$132 & 306$\pm$40 \\
H3 & 1.84 & 2.00 & 2.19 & 2.24 & 458$\pm$98 & 488$\pm$68 \\
H4 & 0.86 & 3.03 & 1.86 & 2.86 & 142$\pm$34 & 326$\pm$44 \\
H5 & 1.13 & 1.44 & 1.96 & 2.96 & 395$\pm$97 & 331$\pm$36 \\
H6 & 0.94 & 2.18 & 2.71 & 4.56 & 216$\pm$77 & 297$\pm$62 \\
\hline
\end{tabular}
\end{adjustbox}
\end{table}

\subsection{$^{14}$N/$^{15}$N ratios derived from HCN and its $^{15}$N isotopologue}

In this section, we test whether the $^{14}$N/$^{15}$N ratios derived in Section 3.1 are significantly affected by $^{13}$C depletion as proposed by the modeling of \citet{Roueff2015}. We have thus carried out direct measurements of the $^{14}$N/$^{15}$N ratios by using the J=1$\rightarrow$0 rotational transitions of HCN and HC$^{15}$N, which were observed simultaneously within our frequency setup. HCN (J=1$\rightarrow$0) is optically thick in IRDC star-forming cores such as the cores in Clouds C and F, or core H1 in Cloud H. Therefore for this test we only use the IRDC cores within our sample that show optically-thin or moderately optically-thick emission (i.e. with $\tau$$\,$$\lesssim$1-2). These cores are G1 and G3 in Cloud G, and H2, H3, H4, and H5 in Cloud H. The rms noise level, integrated intensity, central radial velocity, linewidth, peak intensity and derived optical depth of the HCN (J=1$\rightarrow$0) lines, are shown in Table A.5 in the Appendix. 

Following the same analysis procedures as for H$^{13}$CN in Section 3.1, the $^{14}$N/$^{15}$N ratios were calculated from the column densities of HCN and HC$^{15}$N assuming optically thin emission, T$_{ex}$=15$\,$K, and LTE conditions (see Table 4). For the HC$^{15}$N non-detections, the upper limits to the column density of this molecule were estimated from the 3$\sigma$ rms noise level in the HC$^{15}$N spectra. The derived $^{14}$N/$^{15}$N ratios range from $\geq$67 to $\geq$282. If we compare these values with those from column 6 in Table~\ref{Ratios13C}, we find that the $^{14}$N/$^{15}$N ratios inferred from HCN are systematically lower (by factors 1.2-2.7) with respect to those obtained from H$^{13}$CN. This is in contrast with the results from \citet{Roueff2015} since, from their models, the $^{12}$C/$^{13}$C isotopic ratio measured from HCN should be a factor of $\sim$2 higher than that derived from CN (i.e. at time-scales $\geq$1$\,$Myr for the typical densities of IRDC cores of $\sim$10$^5$cm$^{-3}$; see Figure 4 in their paper). We note that this also holds if the $^{14}$N/$^{15}$N ratios of the moderately-optically thick cores are corrected by their HCN optical depths (with $\tau$(HCN)=0.71-1.84, which corresponds to correction factors of $\sim$1.4-2.2). Except for core H5, the corrected values of the $^{14}$N/$^{15}$N ratios for cores H2, H3 and H4 are, respectively, $\geq$503, 445 and 168,  which are consistent with those inferred from H$^{13}$CN and lie within the uncertainties. Although our sub-sample of optically-thin/moderately-optically thick IRDC cores is small, this test suggests that the $^{14}$N/$^{15}$N ratios obtained using the $^{13}$C containing isotopologues are not strongly affected by $^{13}$C depletion as proposed by the models of \citet{Roueff2015}. In Section 4.5, we discuss the possible reasons for this.}

\begin{table}[!tbp]
\tiny
\caption{Column densities and nitrogen ratios obtained in HCN and its $^{12}$C isotopologue}
\label{RatiosHCN}
\centering
\begin{adjustbox}{width=0.3\textwidth}
\begin{tabular}{cccc}
\hline\hline
 & HCN & HC$^{15}$N & HCN \\
\hline
Core & N$_{\rm tot}$(T$_{\rm ex}$=15K) & N$_{\rm tot}$(T$_{\rm ex}$=15K) & $^{14}$N/$^{15}$N  \\
 & [$\times$10$^{13}$cm$^{-2}$] & [$\times$10$^{11}$cm$^{-2}$] & (T$_{\rm ex}$=15K) \\
\hline
\multicolumn{4}{c}{Cloud G} \\
\hline
G1 & 1.06 & 2.47 & 43$\pm$9 \\
G3 & 1.15 & $\leq$1.72 & $\geq$67$\pm$3 \\
\hline
\multicolumn{4}{c}{Cloud H} \\
\hline
H2 & 3.72 & $\leq$1.32 & $\geq$282$\pm$5 \\
H3 & 5.26 & 2.00 & 263$\pm$49 \\
H4 & 3.65 & 3.03 & 121$\pm$24 \\
H5 & 3.71 & 1.44 & 259$\pm$57 \\
\hline
\end{tabular}
\end{adjustbox}
\end{table}

\section{Discussion}
\subsection{Correlation with star formation activity}
{The chemistry of HCN and HNC is known to be temperature dependent \citep{Pineau1990} and any star formation activity in the core could locally heat the molecular gas enhancing the abundance of HCN (and its isotopologues) over HNC. Therefore, it is important to investigate whether the measured $^{14}$N/$^{15}$N isotopic ratios in HCN and HNC show any correlation with the level of star formation activity in the observed IRDC cores. For this purpose, we adopted the classification of the embedded cores in IRDCs C, F, G and H proposed by \citet{Chambers2009} and \citet{Rathborn2010}. Each of the cores is classified as quiescent, intermediate or active based upon their colour in Spitzer/IRAC 3-8 $\mu$m images as well as the presence or absence of 24 $\mu$m point source emission. The summary of each core's classification is listed in Table~\ref{Cores}. 

In Fig.2, we report the column densities of the $^{15}$N isotopologues against those of the $^{14}$N species in relation to their star-formation classification. This Figure shows that there is no correlation between the column densities of the HCN or HNC isotopologues with the level of star formation activity in the IRDC cores. Such conclusion is also confirmed by plotting the column densities of HC$^{15}$N against that of H$^{15}$NC. In other words, the measurements of $^{14}$N/$^{15}$N ratio towards IRDCs C, F, G and H from the $J$=1$\rightarrow$0 transitions of HCN and HNC indeed probe the chemical composition of the envelope of these IRDCs cores. As such, it is in general not affected by local star formation feedback, although the highest $^{15}$N/$^{14}$N ratio is found towards one of the active cores. Therefore, we cannot rule out that higher-J transitions and higher-angular resolution observations give $^{15}$N/$^{14}$N ratios that are correlated with star formation activity.}

\begin{table*}[!tbp]
\caption{Summary of IRDC cores classification}
\label{Cores}
\centering
\begin{tabular}{cccc}
\hline\hline
IRAC 3-8$\mu$m & 24$\mu$m emission & Core category & Cores in IRDCs C, F, G and H \\
\hline
8.0$\mu$m & Yes/No & Red & - \\
Green Fuzzy & Yes & Active & C2, C3, C6, C9, H2, H5, H6 \\
Green Fuzzy & No & Intermediate & C4, F1 \\
None & Yes & - & - \\
None & No & Quiescent & C1, C5, C8, F2, F3, G2, H1, H3, H4 \\
3.6$\mu$m emission & Yes/No & Blue & - \\
\hline
\multicolumn{4}{l}{Note: Cores F4, G1 and G3 have not yet been classified \citep{Rathborn2010,Chambers2009}.}
\end{tabular}
\end{table*}

\begin{figure}[!tbp]
\centering
\includegraphics[width=0.45\textwidth]{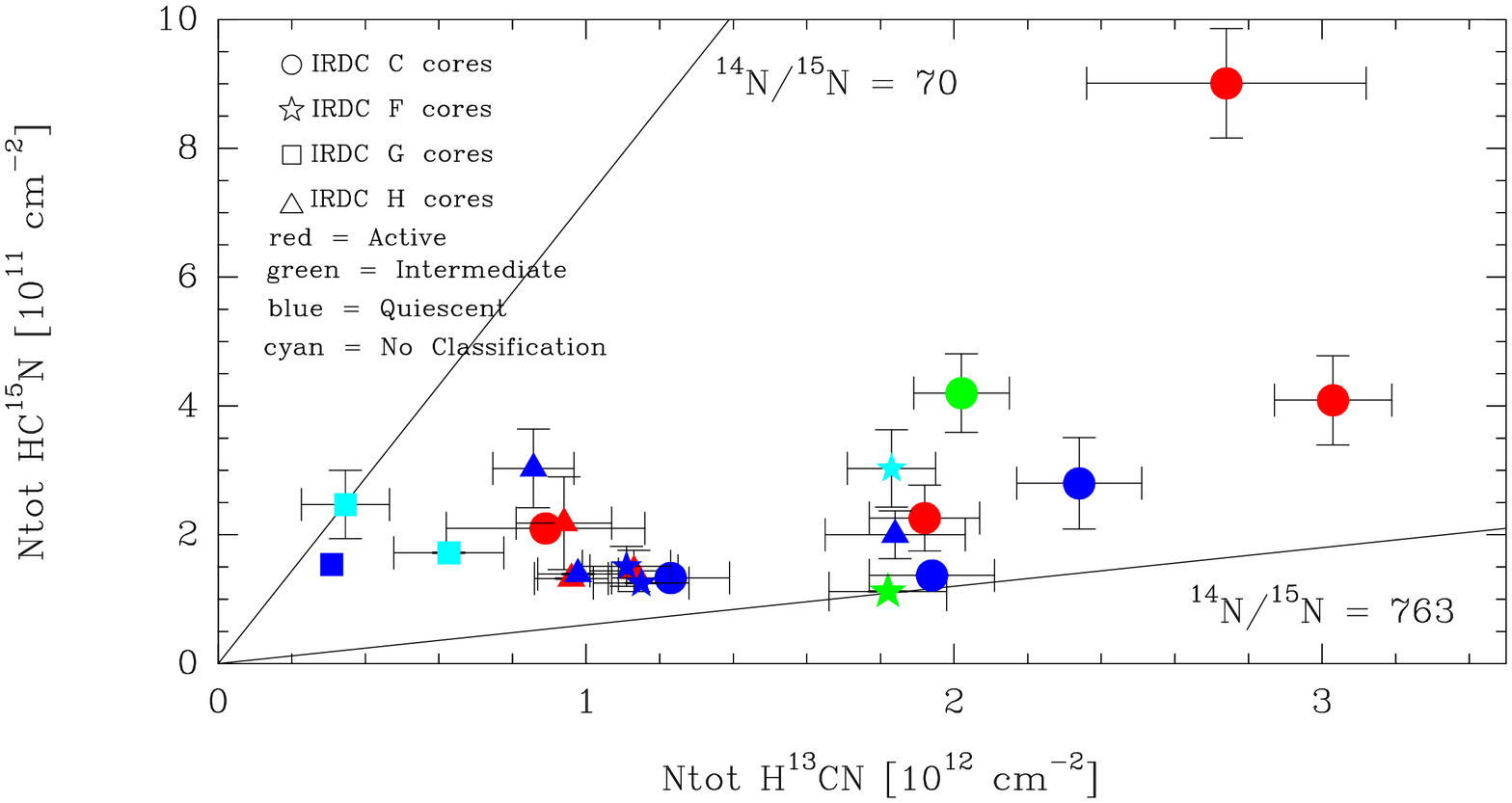}
\includegraphics[width=0.45\textwidth]{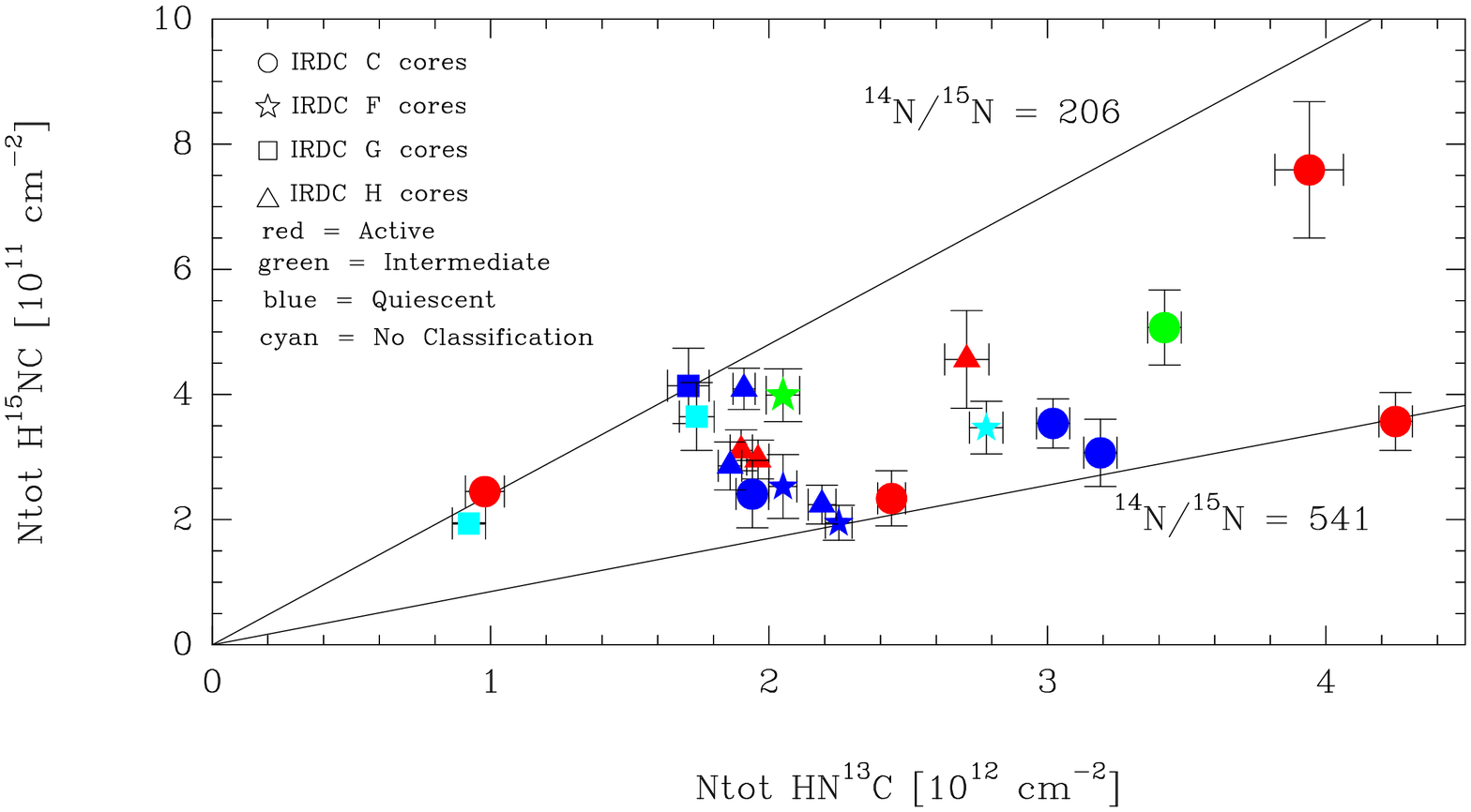}
\caption{\scriptsize Column densities of HCN (top panel) and HNC (bottom panel) $^{15}$N isotopologues plotted against those of the $^{14}$N species for all the cores in the sample. The cores are classified as active (red), intermediate (green), and quiescent (blue). Black indicate cores with no known classification. Different symbols are used to denote different clouds: IRDC C (circle), IRDC F(star), IRDC G (square) and IRDC H (triangle). The straight lines indicate the lowest and highest value of the corresponding $^{14}$N/$^{15}$N ratio for each molecule. }
\centering
\end{figure}

\subsection{Comparison with Solar System objects and low-mass star-forming regions}
{To understand whether IRDC cores have a nitrogen chemical composition consistent with that of the Solar System's birthplace, it is essential to compare the results obtained toward our sample of IRDC cores with those measured in Solar System objects (see Fig 3). For completeness, Fig. 3 also reports the $^{14}$N/$^{15}$N ratios obtained towards low-mass pre-stellar/starless and star-forming cores. Overall, the $^{14}$N/$^{15}$N ratios from Clouds C, F and H show values similar to those observed in the Sun, planets and pre-stellar/star-forming regions, and consistent with the TA and the PSN values. On the other hand, the $^{14}$N/$^{15}$N ratios measured towards Cloud G lie mostly at the level of the TA value and are significantly lower than the PSN value in Fig 3. The results in Cloud G are also in agreement with the measurements obtained in Comets, IDPs, Meteorites and especially with protoplanetary disks (80--160) as recently measured by \citet{Guzman2017}. Furthermore, they marginally agree with the lower end of the ratios derived in starless/pre-stellar and star-forming cores. We also caution that half of the $^{14}$N/$^{15}$N ratios in Cloud G are lower limits (HC$^{15}$N has not been detected in cores G2 and G3; and H$^{15}$NC has not been detected in core G1; see Tables A.2 and A.4), and therefore we may be lacking enough statistics to draw a firm conclusion.}

\subsection{Comparison with high-mass star-forming regions}

{The comparison with the measurements from \citet{Adande2012} and \citet{Fontani2015} towards high-mass star-forming regions, shows that our measurements are consistent with their results as a whole. Especially, the $^{14}$N/$^{15}$N ratios from HNC in IRDCs almost lie in the same range as those measured by \citet{Adande2012} in CN and HNC and by \citet{Fontani2015} in CN. The $^{14}$N/$^{15}$N ratios from HCN are also compatible with the results measured by \citet{Fontani2015} in N$_2$H$^+$ emission. Since \citet{Fontani2015} also measured the $^{14}$N/$^{15}$N ratio towards cores C1, F1, F2 and G2 included in our sample, we have compared the results between each individual core and they show some discrepancies as a result. Indeed, the $^{14}$N/$^{15}$N ratios obtained by \citet{Fontani2015} in N$^{15}$NH$^+$ and $^{15}$NNH$^+$ (CN was not detected towards these cores) are, respectively, $\sim$1445 and $\sim$1217 for C1, $\sim$672 and $\sim$566 for F1 and $\sim$872 and $\sim$856 for G2, i.e. overall significantly higher than those measured in this work. In contrast, F2 shows a lower value of $\sim$232 and $\geq$195 in $^{15}$NNH$^+$ and N$^{15}$NH$^+$ respectively. We note that these large discrepancies have also been found in low-mass pre-stellar cores and could be associated with the different chemistries involved in the formation of N$_2$H$^+$ and HNC/HCN \citep{Wirstrom2012,Hily-Blant2013a,Bizzocchi2013}. More recently, cores C1, C3, F1, F2 and G2 have been studied independently by Colzi et al. (2017, submitted) using isotopologues of HCN and HNC. A similar range for the $^{14}$N/$^{15}$N ratios has been found in HCN ($\geq$150-748) whilst results in HNC lie in a slightly higher range (263-813). In both samples, core G2 shows one of the smallest ratios in HCN and HNC.}

\subsection{$^{13}$C depletion and its effects on nitrogen fractionation}

In Section 3.2, we evaluated whether the depletion of $^{13}$C for species such as HCN and HNC \citep[as predicted by the models of][] {Roueff2015} could affect our derived values of the $^{14}$N/$^{15}$N ratios. Our test revealed that the $^{14}$N/$^{15}$N values inferred from HCN are either consistent, or lower, than those measured from the $^{13}$C isotopologue, in contrast to the modelling predictions. This could be due to two reasons: i) the time-scales (age) of IRDC cores, and ii) the kinetic temperature of the gas within them. 

Regarding the time-scales, \citet{Kong2017} have modelled the chemistry of deuterated species such as N$_2$D$^+$ in IRDC cores to provide constraints to the dynamical age of these cores. Their modelling shows that the enhanced D/H ratio in these objects can be reproduced for time-scales of $\sim$10$^5$$\,$yrs (note that these authors model the N$_2$D$^+$ emission arising from the C1 core in Cloud C). On the other hand, \citet{Roueff2015} predict similar $^{12}$C/$^{13}$C ratios associated with CN and with HCN/HNC at these time-scales (of $\sim$10$^5$$\,$yrs) for the typical H$_2$ gas densities of IRDC cores \citep[of $\sim$10$^5$$,$cm$^{-3}$; see][]{Butler2012}. The large differences in $^{12}$C/$^{13}$C ratios associated with HCN/HNC, such as those discussed in Section 3.2, would therefore not be expected. 

Nevertheless, the definition of a "core" formation timescale is somewhat ambiguous, for example \citet{Barnes2016} found that the D/H fraction within Cloud F would have taken several 10$^{6}$yr to form. In light of this, a more self-consistent comparison between timescales inferred by chemical models is required. 

Concerning the gas temperature of IRDC cores, measurements of the emission of NH$_3$ toward these cores give kinetic temperatures of the gas of 15-20$\,$K \citep{Pillai2006}, which are higher than those assumed in the models of \citet[][of 10$\,$K]{Roueff2015}. Since carbon depletion is strongly dependent on gas/dust temperature, it is unclear whether these results can be compared directly to IRDC cores (note that no models are provided for temperatures higher than 10$\,$K). Therefore, additional modeling is needed to test the effects of $^{13}$C depletion in the chemistry of nitrogen fractionation at slightly higher temperatures similar to those found in IRDCs.}

\subsection{Systematic trend of $^{14}$N/$^{15}$N ratio between IRDCs}

{Table~\ref{Ratios13C} and Figure 3 show that the $^{14}$N/$^{15}$N ratios observed in Cloud G are systematically lower than those measured in Clouds C, F and H.
This may be due to the properties of Cloud G itself. As discussed in Section 4.1, cores G1, G2 and G3 do not show any trace of star-formation activity, whilst the other three IRDCs show several cores that are actively forming stars (see e.g. cores C2 or H2). In addition, Cloud G is the least massive, the most diffuse (it has the weakest emission in high-density tracers; Cosentino et al., in prep.), and it has the lowest peak H$_2$ mass surface density amongst the four targeted IRDCs \citep[see Table 2 and][]{Butler2012}. Given that the kinetic temperature of the gas is similar across IRDCs ($\sim$15-20$\,$K), we propose that density could be one of the important parameters that is responsible for the discrepancies found between Cloud G and the other IRDCs, although the models do not agree with this scenario. Therefore, we speculate that the PSN may have formed in an IRDC with properties similar to those of cloud G. However, we note that the properties of this sample of IRDC cores need to be further investigated along with relevant chemical models in order to confirm the proposed idea.}

\begin{figure}[!tbp]
\includegraphics[width=0.45\textwidth]{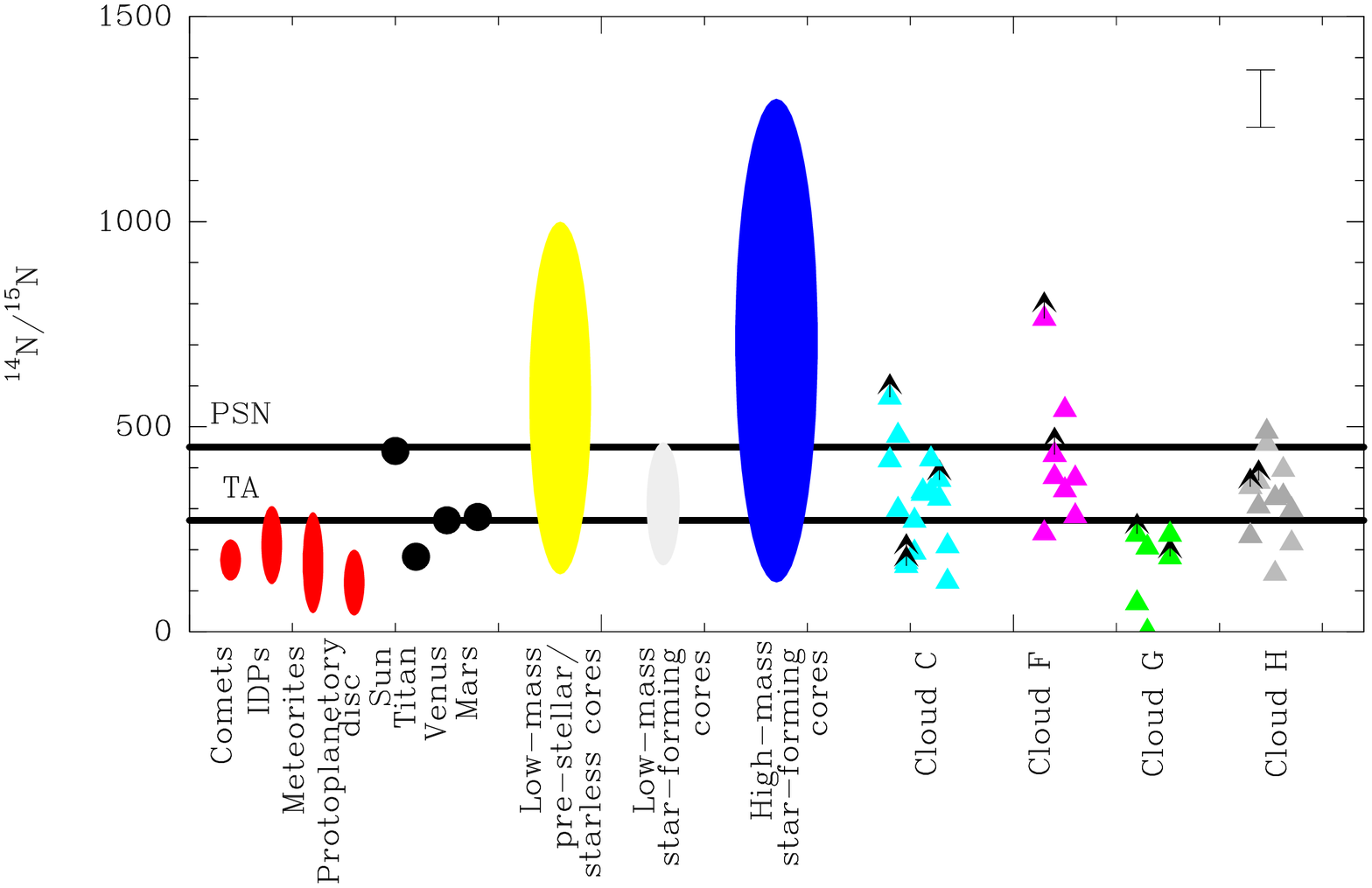}
\caption{\scriptsize Nitrogen isotope ratio variations measured in different sources starting from small Solar System bodies \citep[in red;][]{Busemann2006,Floss2006,Alexander2007,Bockelee-Morvan2008,Manfroid2009,Mumma2011,Guzman2017} to planets \citep[in black;][]{Junk1958,Hoffman1979,Fouchet2004,Niemann2005,Marty2010,Wong2013,Fletcher2014}, low-mass pre-stellar cores  \citep[in yellow;][]{Gerin2009,Bizzocchi2010,Lis2010,Bizzocchi2013,Hily-Blant2013a}, low-mass star-forming cores \citep[in light grey;][]{Wampfler2014} and high-mass star-forming cores \citep[in blue;][]{Adande2012,Fontani2015}. Our measurements in IRDCs are shown on the right. On the upper right corner, we show a representative error bar for our measurements of the $^{14}$N/$^{15}$N ratio. Black arrows indicate the lower limits in our measurements. Note that the $^{14}$N/$^{15}$N ratio measured in HCN from core G2 is shown as zero with no lower limit indication due to both of the column densities of H$^{13}$CN and HC$^{15}$N are only given as upper limits.}
\centering
\end{figure}

\section{Conclusions}

{We have measured the nitrogen isotopic ratio $^{14}$N/$^{15}$N toward a sample of cold IRDC cores. This ratio ranges between $\sim$70 and $\geq$763 in HCN and between $\sim$161 and $\sim$541 in HNC. In particular, Cloud G systematically shows lower nitrogen isotopic ratios than the other three clouds, with values being consistent with the ratio measured toward small Solar System bodies such as Comets, IDPs, Meteorites and also proto-planetary disks. Since Cloud G shows lower overall gas densities, and since it likely is at an earliest stage of evolution, we propose that gas density is the key parameter in nitrogen fractionation in IRDCs. Higher angular resolution observations, as well as chemical modeling of the nitrogen fractionation of HCN and HNC at temperatures similar to those found in IRDCs, are needed to establish the origin of the discrepancies in the measured $^{14}$N/$^{15}$N ratios found in Cloud G with respect to Clouds C, F and H. The comparison between the modeling predictions and our IRDC measurements may allow to constrain the main chemical reactions involved in the fractionation process of Nitrogen in the proto-solar nebula.}

\begin{acknowledgements}
We would like to thank an anonymous referee for the valuable comments to a previous version of the manuscript. I.J.-S. acknowledges the financial support received from the STFC through an Ernest Rutherford Fellowship (proposal number ST/L004801/2). P.C. acknowledges financial support of the European Research Council (ERC; project PALs 320620). The research leading to these results has also received funding from the European Commission (FP/2007-2013) under grant agreement No. 283393 (RadioNet3)."
\end{acknowledgements}

\bibliographystyle{aa}
\bibliography{ref}

\appendix

\section{Spectra and fitting parameters of isotopologues of HCN and HNC}
\begin{figure*}
\centering
\includegraphics[width=\textwidth]{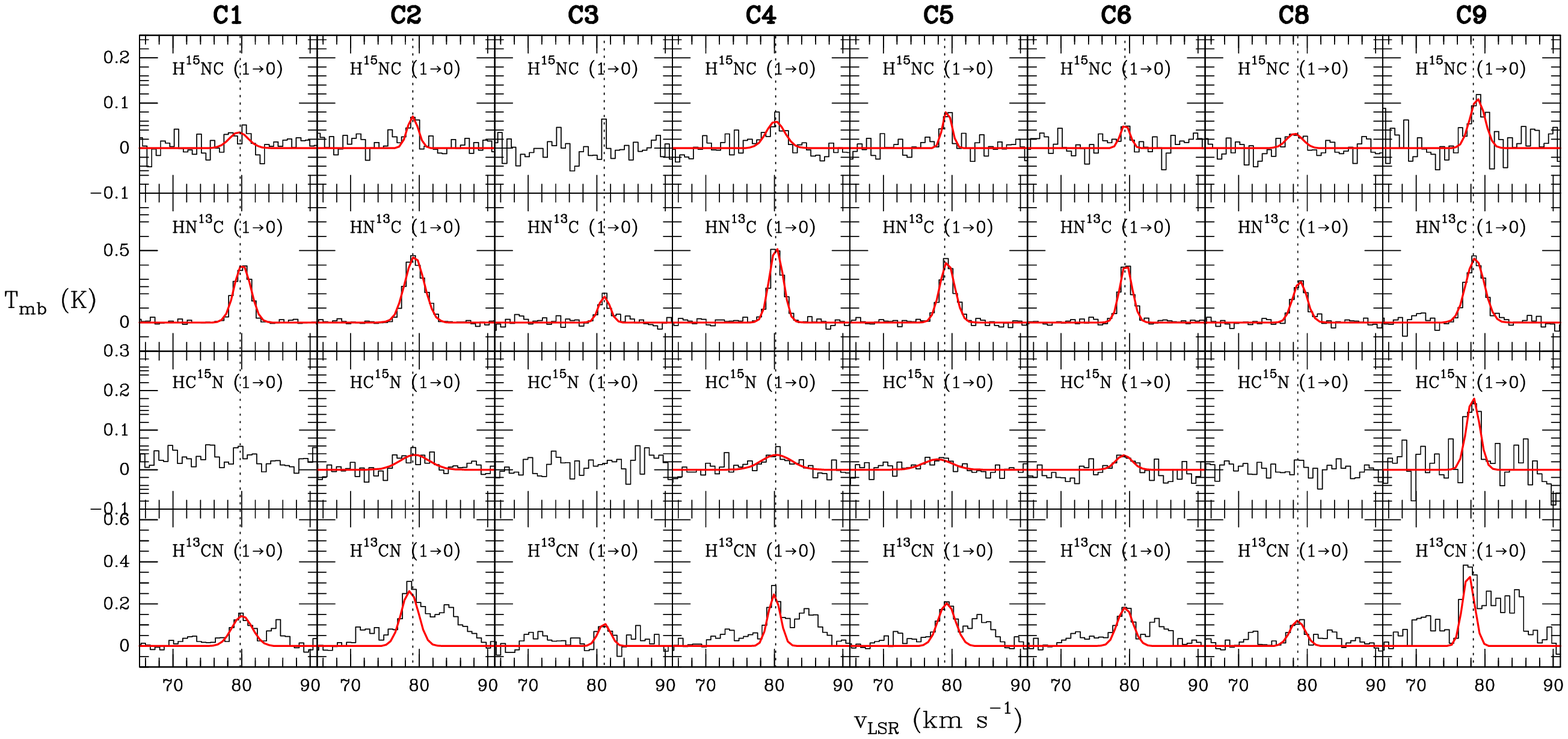}
\caption{Spectra of HC$^{15}$N, HN$^{13}$C H$^{15}$NC and H$^{13}$CN observed with IRAM-30m toward IRDC C. The red line indicates the best Gaussian fit. Note that the hyperfine components of H$^{13}$CN were initially fitted but given the bad results of the fit, in a second step only the main $F$=2$\rightarrow$1 component of H$^{13}$CN was fitted using a single-Gaussian component profile (see red line in bottom panels and Section 3.1 for details).}
\end{figure*}

\begin{figure*}
\centering
\includegraphics[width=0.6\textwidth]{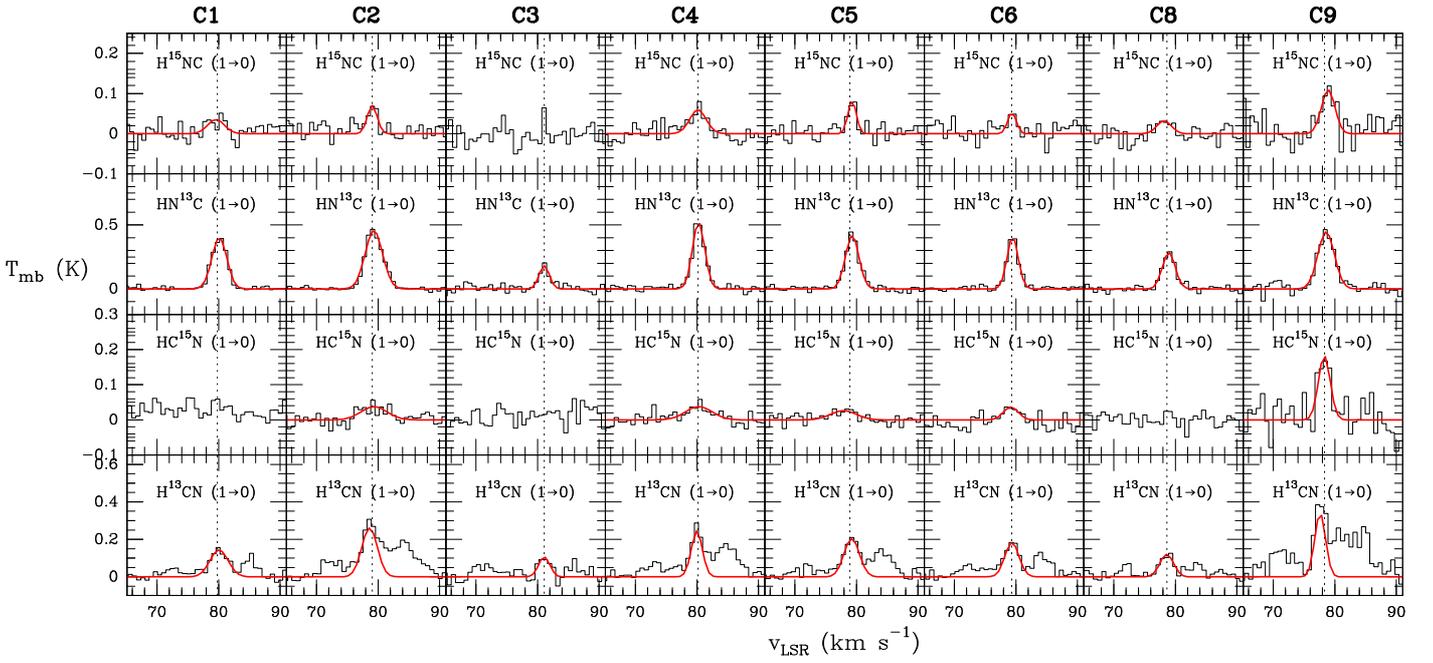}
\caption{Spectra of HC$^{15}$N, HN$^{13}$C H$^{15}$NC and H$^{13}$CN observed with IRAM-30m toward IRDC F. The red line indicates the best Gaussian fit. Note that the hyperfine components of H$^{13}$CN were initially fitted but given the bad results of the fit, in a second step only the main $F$=2$\rightarrow$1 component of H$^{13}$CN was fitted using a single-Gaussian component profile (see red line in bottom panels and Section 3.1 for details).}
\end{figure*}

\begin{figure*}
\centering
\includegraphics[width=0.45\textwidth]{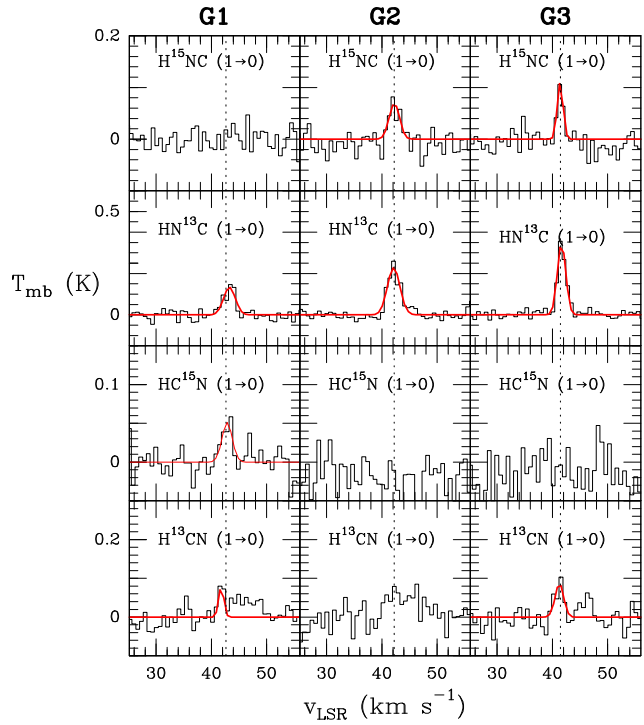}
\caption{Spectra of HC$^{15}$N, HN$^{13}$C H$^{15}$NC and H$^{13}$CN observed with IRAM-30m toward IRDC G. Note that only the main component of H$^{13}$CN was fitted with a single-Gaussian component profile, which was then used to calculate the column densities. The red line indicates the best Gaussian fit and these spectra have been scaled to fit into each individual panel.}
\end{figure*}

\begin{figure*}
\centering
\includegraphics[width=0.8\textwidth]{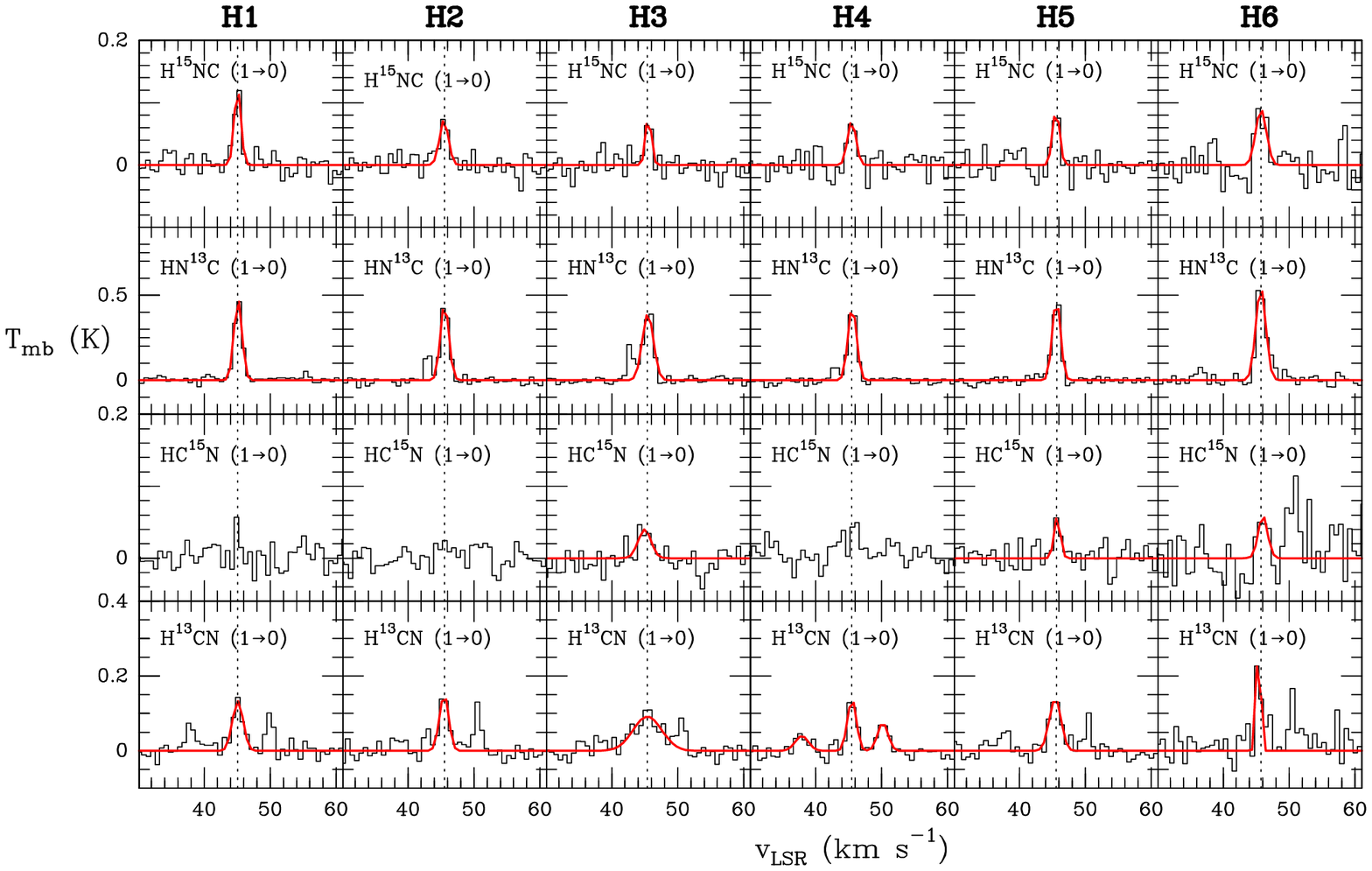}
\caption{Spectra of HC$^{15}$N, HN$^{13}$C H$^{15}$NC and H$^{13}$CN observed with IRAM-30m toward IRDC H. The red line indicates the best Gaussian fit. Note that the hyperfine components of H$^{13}$CN were initially fitted but given the bad results of the fit, in a second step only the main $F$=2$\rightarrow$1 component of H$^{13}$CN was fitted using a single-Gaussian component profile (see red line in bottom panels and Section 3.1 for details).}
\end{figure*}

\begin{table*}[!tbp]
\tiny
\caption{The root mean square noise in the spectra rms, integrated intensities $\int$ T$_{\rm mb}$d$\upsilon$, radial velocity $\upsilon$, line widths $\Delta$$\upsilon$ and peak temperature T$_{\rm peak}$ of HC$^{15}$N(1$\rightarrow$0) in cores within Clouds C, F, G and H.} 
\centering
\begin{tabular}{lccccc} 
\hline\hline
Core & RMS [10$^{-2}$ K] & $\int$ T$_{\rm mb}$$\Delta$$\upsilon$ [K$\,$kms$^{-1}$] & $\upsilon$ [kms$^{-1}$] & $\Delta$$\upsilon$ [kms$^{-1}$] & T$_{\rm peak}$ [K] \\ 
\hline
\multicolumn{6}{c}{Cloud C} \\
\hline
C1* & 1.8 & $\leq$0.07 & - & - & $\leq$0.05 \\
C2 & 1.8 & 0.2$\pm$0.05 & 79.36$\pm$0.62 & 5.02$\pm$1.07 & 0.04 \\
C3* & 2.7 & $\leq$0.10 & - & - & $\leq$0.08 \\
C4 & 1.6 & 0.21$\pm$0.05 & 80.31$\pm$0.59 & 5.12$\pm$1.71 & 0.04 \\
C5 & 1.9 & 0.14$\pm$0.05 & 77.89$\pm$0.89 & 4.90$\pm$2.16 & 0.03 \\
C6 & 1.8 & 0.11$\pm$0.04 & 79.19$\pm$0.50 & 2.81$\pm$1.01 & 0.04 \\
C7 & - & - & - & - & - \\
C8* & 1.7 & $\leq$0.06 & - & - & $\leq$0.05 \\
C9 & 3.3 & 0.44$\pm$0.07 & 78.30$\pm$0.17 & 2.26$\pm$0.37 & 0.18 \\
\hline\multicolumn{6}{c}{Cloud F} \\
\hline
F1* & 1.5 & $\leq$0.05 & - & - & $\leq$0.04 \\
F2* & 1.6 & $\leq$0.06 & - & - & $\leq$0.05 \\
F3 & 1.5 & 0.07$\pm$0.02 & 57.9$\pm$0.22 & 1.49$\pm$0.49 & 0.05 \\
F4 & 1.8 & 0.15$\pm$0.04 & 57.74$\pm$0.54 & 3.93$\pm$1.05 & 0.04 \\
\hline\multicolumn{6}{c}{Cloud G} \\
\hline
G1 & 2.1 & 0.12$\pm$0.04 & 42.85$\pm$0.35 & 2.20$\pm$0.67 & 0.05 \\
G2* & 2.0 & $\leq$0.08 & - & - & $\leq$0.06 \\
G3* & 2.2 & $\leq$0.08 & - & - & $\leq$0.07 \\
\hline\multicolumn{6}{c}{Cloud H} \\
\hline
H1* & 1.8 & $\leq$0.07 & - & - & $\leq$0.05 \\
H2* & 1.7 & $\leq$0.06 & - & - & $\leq$0.05 \\
H3 &  1.5 & 0.10$\pm$0.03 & 44.97$\pm$0.33 & 2.28$\pm$0.86 & 0.04 \\
H4 & 1.8 & 0.15$\pm$0.04 & 45.37$\pm$0.56 & 3.99$\pm$1.17 & 0.03 \\
H5 & 1.8 & 0.07$\pm$0.02 & 45.74$\pm$0.19 & 1.13$\pm$0.48 & 0.06 \\
H6 & 3.3 & 0.11$\pm$0.05 & 46.03$\pm$0.43 & 1.68$\pm$0.63 & 0.06 \\
\hline
\multicolumn{6}{l}{(a) * indicates 3$\sigma$ upper-limit values for T$_{peak}$.}\\
\multicolumn{6}{l}{(b) The integrated intensity upper limits are calculated as 3$\sigma \times$$\sqrt{\delta v \times \Delta v}$, with $\delta v$ the spectral} \\
\multicolumn{6}{l}{resolution (0.68 km s$^{-1}$) and $\Delta v$ the average linewidth measured considering all cores.}
\end{tabular}
\end{table*}

\begin{table*}[!tbp]
\tiny
\caption{The root mean square noise in the spectra rms, integrated intensities $\int$ T$_{\rm mb}$d$\upsilon$, radial velocity $\upsilon$, line widths $\Delta$$\upsilon$ and peak temperature T$_{\rm peak}$ of HN$^{13}$C(1$\rightarrow$0) in cores within Clouds C, F, G and H.} 
\centering
\begin{tabular}{cccccc} 
\hline\hline
Core & RMS [10$^{-2}$ K] & $\int$ T$_{\rm mb}$$\Delta$$\upsilon$ [K$\,$kms$^{-1}$] & $\upsilon$ [kms$^{-1}$] & $\Delta$$\upsilon$ [kms$^{-1}$] & T$_{\rm peak}$ [K] \\ 
\hline
\multicolumn{6}{c}{Cloud C} \\
\hline
C1 & 1.8 & 1.22$\pm$0.03 & 80.12$\pm$0.04 & 2.84$\pm$0.09 & 0.41 \\
C2 & 1.7 & 1.62$\pm$0.04 & 79.32$\pm$0.04 & 3.32$\pm$0.08 & 0.46 \\
C3 & 2.3 & 0.37$\pm$0.05 & 81.09$\pm$0.11 & 1.93$\pm$0.34 & 0.18 \\
C4 & 1.6 & 1.30$\pm$0.03 & 80.26$\pm$0.03 & 2.35$\pm$0.07 & 0.52 \\
C5 & 1.6 & 1.15$\pm$0.04 & 79.31$\pm$0.04 & 2.57$\pm$0.10 & 0.42 \\
C6 & 1.7 & 0.93$\pm$0.03 & 79.47$\pm$0.04 & 2.15$\pm$0.09 & 0.41 \\
C7 & - & - & - & - & - \\
C8 & 1.7 & 0.74$\pm$0.04 & 78.94$\pm$0.06 & 2.44$\pm$0.14 & 0.29 \\
C9 & 3.2 & 1.50$\pm$0.07 & 78.59$\pm$0.07 & 3.16$\pm$0.17 & 0.45 \\
\hline\multicolumn{6}{c}{Cloud F} \\
\hline
F1 & 1.5 & 0.78$\pm$0.03 & 57.61$\pm$0.07  &3.39$\pm$0.15 & 0.22 \\
F2 & 1.6 & 0.78$\pm$0.03 & 58.09$\pm$0.04 & 2.22$\pm$0.10 & 0.33 \\
F3 & 1.5 & 0.86$\pm$0.03 & 58.25$\pm$0.04 & 2.12$\pm$0.09 & 0.38 \\
F4 & 1.7 & 1.06$\pm$0.03 & 58.85$\pm$0.04 & 2.35$\pm$0.08 & 0.42 \\
\hline\multicolumn{6}{c}{Cloud G} \\
\hline
G1 & 1.9 & 0.35$\pm$0.04 & 43.28$\pm$0.15 & 2.47$\pm$0.34 & 0.13 \\
G2 & 2.1 & 0.65$\pm$0.04 & 42.15$\pm$0.08 & 2.67$\pm$0.19 & 0.23 \\
G3 & 2.2 & 0.66$\pm$0.03 & 41.68$\pm$0.05 & 1.79$\pm$0.11 & 0.35 \\
\hline\multicolumn{6}{c}{Cloud H} \\
\hline
H1 & 1.7 & 0.73$\pm$0.03 & 45.13$\pm$0.02 & 1.43$\pm$0.06 & 0.48 \\
H2 & 1.6 & 0.73$\pm$0.04 & 45.57$\pm$0.03 & 1.50$\pm$0.09 & 0.46 \\
H3 & 1.6 & 0.84$\pm$0.05 & 45.48$\pm$0.06 & 2.00$\pm$0.16 & 0.39 \\
H4 & 1.6 & 0.71$\pm$0.03 & 45.58$\pm$0.03 & 1.53$\pm$0.07 & 0.44 \\
H5 & 1.7 & 0.75$\pm$0.03 & 45.64$\pm$0.03 & 1.46$\pm$0.06 & 0.48 \\
H6 & 2.8 & 1.03$\pm$0.05 & 45.73$\pm$0.04 & 1.77$\pm$0.09 & 0.55 \\
\hline

\multicolumn{6}{l}{(a) The integrated intensity upper limits are calculated as 3$\sigma \times$$\sqrt{\delta v \times \Delta v}$, with $\delta v$ the spectral} \\
\multicolumn{6}{l}{resolution (0.68 km s$^{-1}$) and $\Delta v$ the average linewidth measured considering all cores.}
\end{tabular}
\end{table*}

\begin{table*}[!tbp]
\tiny
\caption{The root mean square noise in the spectra rms, integrated intensities $\int$ T$_{\rm mb}$d$\upsilon$, radial velocity $\upsilon$, line widths $\Delta$$\upsilon$ and peak temperature T$_{\rm peak}$ of H$^{15}$NC(1$\rightarrow$0) in cores within Clouds C, F, G and H.} 
\centering
\begin{tabular}{cccccc} 
\hline\hline
Core & RMS [10$^{-2}$ K] & $\int$ T$_{\rm mb}$$\Delta$$\upsilon$ [K$\,$kms$^{-1}$] & $\upsilon$ [kms$^{-1}$] & $\Delta$$\upsilon$ [kms$^{-1}$] & T$_{\rm peak}$ [K] \\ 
\hline
\multicolumn{6}{c}{Cloud C} \\
\hline
C1 & 1.4 & 0.12$\pm$0.03 & 79.58$\pm$0.47 & 3.24$\pm$0.72 & 0.04 \\
C2 & 1.6 & 0.14$\pm$0.03 & 79.06$\pm$0.20 & 1.88$\pm$0.57 & 0.07 \\
C3* & 2.6 & $\leq$0.10 & - & - & $\leq$0.07 \\
C4 & 1.5 & 0.20$\pm$0.04 & 80.16$\pm$0.30 & 3.17$\pm$0.85 & 0.06 \\
C5 & 1.5 & 0.14$\pm$0.02 & 79.30$\pm$0.14 & 1.57$\pm$0.32 & 0.08 \\
C6 & 1.7 & 0.09$\pm$0.03 & 79.37$\pm$0.27 & 1.66$\pm$0.79 & 0.05 \\
C7 & - & - & - & - & - \\
C8 & 1.5 & 0.10$\pm$0.03 & 78.17$\pm$0.47 & 2.76$\pm$0.83 & 0.03 \\
C9 & 3.3 & 0.30$\pm$0.07 & 78.92$\pm$0.27 & 2.58$\pm$0.07 & 0.11 \\
\hline\multicolumn{6}{c}{Cloud F} \\
\hline
F1 & 0.9 & 0.09$\pm$0.02 & 56.14$\pm$0.13 & 1.21$\pm$0.30 & 0.03 \\
F2 & 1.6 & 0.10$\pm$0.03 & 57.92$\pm$0.42 & 2.50$\pm$1.12 & 0.04 \\
F3 & 1.3 & 0.08$\pm$0.02 & 57.90$\pm$0.12 & 1.00$\pm$0.21 & 0.07 \\
F4 & 1.5 & 0.14$\pm$0.03 & 58.95$\pm$0.32 & 2.69$\pm$0.61 & 0.05 \\
\hline\multicolumn{6}{c}{Cloud G} \\
\hline
G1* & 2.0 & $\leq$0.08 & - & - & $\leq$0.06 \\
G2 & 1.9 & 0.16$\pm$0.03 & 42.26$\pm$0.26 & 2.22$\pm$0.46 & 0.07 \\
G3 & 2.3 & 0.14$\pm$0.03 & 41.42$\pm$0.13 & 1.24$\pm$0.26 & 0.11 \\
\hline\multicolumn{6}{c}{Cloud H} \\
\hline
H1 & 1.5 & 0.16$\pm$0.02 & 45.05$\pm$0.07 & 1.20$\pm$0.22 & 0.13 \\
H2 & 1.5 & 0.12$\pm$0.03 & 45.44$\pm$0.15 & 1.60$\pm$0.51 & 0.07 \\
H3 & 1.4 & 0.09$\pm$0.02 & 45.56$\pm$0.12 & 1.04$\pm$0.37 & 0.08 \\
H4 & 1.5 & 0.11$\pm$0.02 & 45.49$\pm$0.15 & 1.50$\pm$0.21 & 0.07 \\
H5 & 1.4 & 0.12$\pm$0.02 & 45.57$\pm$0.11 & 1.23$\pm$0.26 & 0.09 \\
H6 & 2.7 & 0.18$\pm$0.04 & 45.75$\pm$0.24 & 1.90$\pm$0.38 & 0.09 \\
\hline
\multicolumn{6}{l}{(a) * indicates 3$\sigma$ upper-limit values for T$_{peak}$.}\\
\multicolumn{6}{l}{(b) The integrated intensity upper limits are calculated as 3$\sigma \times$$\sqrt{\delta v \times \Delta v}$, with $\delta v$ the spectral} \\
\multicolumn{6}{l}{resolution (0.68 km s$^{-1}$) and $\Delta v$ the average linewidth measured considering all cores.}
\end{tabular}
\end{table*}

\begin{table*}[!tbp]
\tiny
\caption{The root mean square noise in the spectra rms, integrated intensities $\int$ T$_{\rm mb}$d$\upsilon$, radial velocity $\upsilon$, line widths $\Delta$$\upsilon$, peak temperature T$_{\rm peak}$ of the H$^{13}$CN (1$\rightarrow$0, $F$=2$\rightarrow$1) hyperfine component and optical depth $\tau$ of the H$^{13}$CN(1$\rightarrow$0) emission in cores within Clouds C, F, G and H.} 
\centering
\begin{tabular}{ccccccc} 
\hline\hline
Core & RMS [10$^{-2}$ K] & $\int$ T$_{\rm mb}$$\Delta$$\upsilon$ [K$\,$kms$^{-1}$] & $\upsilon$ [kms$^{-1}$] & $\Delta$$\upsilon$ [kms$^{-1}$] & T$_{\rm peak}$ [K] & $\tau$ \\ 
\hline
\multicolumn{7}{c}{Cloud C} \\
\hline
C1 & 1.8 & 0.53$\pm$0.05 & 80.08$\pm$0.14 & 3.46$\pm$0.39 & 0.14 & 1.47$\pm$1.18 \\
C2 & 1.8 & 0.83$\pm$0.13 & 78.63$\pm$0.10 & 2.97$\pm$0.31 & 0.26 & 1.79$\pm$0.57 \\
C3 & 3.4 & 0.24$\pm$0.06 & 81.08$\pm$0.27 & 2.13$\pm$0.46 & 0.11 & 2.86$\pm$4.00 \\
C4 & 1.6 & 0.55$\pm$0.07 & 79.98$\pm$0.07 & 2.11$\pm$0.22 & 0.25 & 1.75$\pm$0.63 \\
C5 & 1.9 & 0.63$\pm$0.07 & 79.24$\pm$0.13 & 2.95$\pm$0.35 & 0.20 & 1.33$\pm$0.81 \\
C6 & 1.7 & 0.52$\pm$0.07 & 79.36$\pm$0.12 & 2.69$\pm$0.30 & 0.18 & 1.87$\pm$1.00 \\
C7 & - & - & - & - & - & - \\
C8 & 1.8 & 0.33$\pm$0.04 & 78.60$\pm$0.15 & 2.61$\pm$0.39 & 0.12 & 2.11$\pm$2.09 \\
C9 & 4.8 & 0/75$\pm$0.15 & 77.62$\pm$0.13 & 2.05$\pm$0.29 & 0.34 & 5.88$\pm$2.22 \\
\hline\multicolumn{6}{c}{Cloud F} \\
\hline
F1 & 1.5 & 0.50$\pm$0.04 & 57.77$\pm$0.14 & 3.57$\pm$0.40 & 0.13 & 2.29$\pm$1.25 \\
F2 & 1.6 & 0.31$\pm$0.03 & 58.04$\pm$0.11 & 2.36$\pm$0.26 & 0.13 & 1.17$\pm$1.43 \\
F3 & 1.5 & 0.30$\pm$0.03 & 58.21$\pm$0.08 & 1.95$\pm$0.18 & 0.15 & 1.32$\pm$1.36 \\
F4 & 1.5 & 0.50$\pm$0.05 & 58.81$\pm$0.07 & 2.36$\pm$0.18 & 0.20 & 0.81$\pm$0.66 \\
\hline\multicolumn{6}{c}{Cloud G} \\
\hline
G1 & 2.1 & 0.09$\pm$0.03 & 41.79$\pm$0.16 & 1.11$\pm$0.36 & 0.08 & 5.35$\pm$5.78 \\
G2* & 2.2 & $\leq$0.08 & - & - & $\leq$0.07 & 0.1$\pm$2.34 \\
G3 & 2.0 & 0.17$\pm$0.03 & 41.27$\pm$0.20 & 1.84$\pm$0.39 & 0.09 & 1.25$\pm$3.04 \\
\hline\multicolumn{6}{c}{Cloud H} \\
\hline
H1 & 1.5 & 0.27$\pm$0.03 & 45.09$\pm$0.10 & 1.89$\pm$0.28 & 0.13 & 4.87$\pm$2.44 \\
H2 & 1.5 & 0.26$\pm$0.03 & 45.47$\pm$0.08 & 1.62$\pm$0.32 & 0.15 & 1.38$\pm$1.44 \\
H3 & 1.5 & 0.50$\pm$0.05 & 45.41$\pm$0.23 & 5.13$\pm$0.63 & 0.09 & 0.10$\pm$3.98 \\
H4 & 1.6 & 0.23$\pm$0.02 & 45.51$\pm$0.08 & 1.57$\pm$0.18 & 0.14 & 0.81$\pm$1.43 \\
H5 & 1.5 & 0.31$\pm$0.03 & 45.43$\pm$0.10 & 2.06$\pm$0.24 & 0.14 & 0.56$\pm$1.69 \\
H6 & 2.8 & 0.26$\pm$0.03 & 45.37$\pm$0.06 & 0.74$\pm$0.22 & 0.33 & 0.10$\pm$1.58 \\
\hline
\multicolumn{6}{l}{(a) * indicates 3$\sigma$ upper-limit values for T$_{peak}$.}\\
\multicolumn{6}{l}{(b) The integrated intensity upper limits are calculated as 3$\sigma \times$$\sqrt{\delta v \times \Delta v}$, with $\delta v$ the spectral} \\
\multicolumn{6}{l}{resolution (0.68 km s$^{-1}$) and $\Delta v$ the average linewidth measured considering all cores.}
\end{tabular}
\end{table*}

\begin{table*}[!tbp]
\tiny
\caption{The root mean square noise in the spectra rms, integrated intensities $\int$ T$_{\rm mb}$d$\upsilon$, radial velocity $\upsilon$, line widths $\Delta$$\upsilon$, peak temperature T$_{\rm peak}$ of the HCN (1$\rightarrow$0, $F$=2$\rightarrow$1) hyperfine component and optical depth $\tau$ of the HCN(1$\rightarrow$0) emission in cores within Clouds G and H.} 
\centering
\begin{tabular}{ccccccc} 
\hline\hline
Core & rms [10$^{-2}$ K] & $\int$ T$_{\rm mb}$$\Delta$$\upsilon$ [K$\,$kms$^{-1}$] & $\upsilon$ [kms$^{-1}$] & $\Delta$$\upsilon$ [kms$^{-1}$] & T$_{\rm peak}$ [K] & $\tau$ \\ 
\hline
\multicolumn{6}{c}{Cloud G} \\
\hline
G1 & 1.4 & 0.95$\pm$0.07 & 42.66$\pm$0.16 & 4.05$\pm$0.31 & 0.22 & 0.10$\pm$0.16 \\
G3 & 2.3 & 1.03$\pm$0.05 & 40.43$\pm$0.05 & 1.17$\pm$0.09 & 0.82 & 0.10$\pm$0.02 \\
\hline\multicolumn{6}{c}{Cloud H} \\
\hline
H2 & 1.6 & 3.32$\pm$0.05 & 44.06$\pm$0.02 & 3.19$\pm$0.05 & 0.98 & 1.29$\pm$0.18 \\
H3 & 1.4 & 4.71$\pm$0.06 & 43.80$\pm$0.02 & 3.48$\pm$0.05 & 1.27 & 1.16$\pm$0.13 \\
H4 & 1.5 & 3.26$\pm$0.06 & 44.47$\pm$0.03 & 3.34$\pm$0.07 & 0.92 & 0.71$\pm$0.04 \\
H5 & 2.0 & 3.32$\pm$0.05 & 44.43$\pm$0.04 & 3.36$\pm$0.06 & 0.93 & 1.84$\pm$0.29 \\
\hline
\end{tabular}
\end{table*}

%

\end{document}